\documentclass[pra,twocolumn,showpacs,preprintnumbers,floatfix]{revtex4-1}
\usepackage{graphicx}
\usepackage{dcolumn}
\usepackage{bm}
\usepackage{latexsym,epsfig}
\usepackage{graphicx}
\usepackage{verbatim}
\usepackage{comment}
\usepackage{amsmath}
\usepackage{amssymb}
\usepackage{stmaryrd}
\usepackage{color}
\usepackage{float}
\restylefloat{table}

\newcommand{\beq}{\begin{equation}}
\newcommand{\eeq}{\end{equation}}
\newcommand{\bea}{\begin{eqnarray}}
\newcommand{\eea}{\end{eqnarray}}
\newcommand{\ben}{\begin{eqnarray*}}
\newcommand{\een}{\end{eqnarray*}}
\newcommand{\bfig}{\begin{figure}}
\newcommand{\efig}{\end{figure}}

\begin{document}
\title{Permanent Electric Dipole Moments of Alkaline Earth Monofluorides: Interplay of Relativistic and Correlation Effects}
\author{V S Prasannaa$^1,^2$, Sreerekha S$^2$, M Abe$^3,^4$, V M Bannur$^2$, B P Das$^5$}
\affiliation{$^1$Indian Institute of Astrophysics, II Block, Koramangala, Bangalore-560 034, India}
\affiliation{$^2$Department of Physics, Calicut University, Malappuram, Kerala-673 635, India}
\affiliation{$^3$Tokyo Metropolitan University, 1-1, Minami-Osawa, Hachioji-city, Tokyo 192-0397, Japan}
\affiliation{$^4$ JST, CREST, 4-1-8 Honcho, Kawaguchi, Saitama 332-0012, Japan}
\affiliation{$^5$ International Education and Research Center of Science (IERCS) and Department of Physics, Tokyo Institute of Technology, 2-12-1-H86 Ookayama,
Meguro-ku, Tokyo  152-8550, Japan}

\date{\today}

\begin{abstract}
The interplay of the relativistic and correlation effects in the permanent electric dipole moments (PDMs) of the $X^2\Sigma^+$ ($\nu=0$) electronic 
ground states of the alkaline earth monoflourides (BeF, MgF, CaF, SrF and BaF) has been studied using a relativistic coupled cluster method (RCCM). 
The calculations were carried out using double, 
triple and quadruple zeta basis sets, and with no core orbitals frozen. The results are 
compared with those of other calculations available in the  literature and with experiments. The correlation trends in the PDMs of these molecules are 
discussed in detail. 

\end{abstract}

\pacs{75.40.Gb, 67.85.-d, 71.27.+a }

\maketitle

\section{Introduction}
In this work, we report the results of our calculations of the permanent electric dipole moments (PDMs) of the alkaline earth monofluorides
(BeF, MgF, CaF, SrF and BaF)  using a RCCM, and elucidate the trends in the correlation 
effects in the PDMs, as the molecules become progressively heavier, and consequently the relativistic effects get more pronounced. To the best of our 
knowledge, 
such a study involving the interplay of relativistic and correlation effects in molecules has not been performed earlier.
The coupled cluster method is considered to be the current gold standard of electronic structure calculations for atoms and diatomic molecules~\cite{Eliav}. 
It is suitable for calculating a number of static properties, including the PDMs of diatomic molecules~\cite{VSP}. The reasons  
for the choice of PDMs of the alkaline earth 
monofluorides are that this property has been calculated earlier using different approaches 
for this class of molecules~\cite{Torring,Rice,Langhoff,Mestdagh,Bundgen,Allouche,Buckingham,Kobus,VSP,Sudip} 
and high precision experimental data are available for some of these 
molecules~\cite{Childs,Ernst,Kandler}. 
Moreover, PDMs are 
important for various applications, especially SrF, which was the first molecule to be laser cooled~\cite{Shuman}, and for BaF, which is 
a promising candidate for probing fundamental physics~\cite{demille,Mal}. 

\section{Theory}
We discuss briefly the underlying ideas of the PDM of a molecule and the coupled cluster method. The details of both of these topics are discussed 
in detail elsewhere~\cite{VSP}. 
The PDM of a molecule, d, is given by:

\begin{eqnarray}
 d &=& \frac{\langle \psi \arrowvert D \arrowvert \psi \rangle}{\langle \psi \arrowvert \psi \rangle} \nonumber \\
 &=& \langle \Phi_0 \arrowvert e^{T \dag} D_N e^T \arrowvert \Phi_0 \rangle_C + \langle \Phi_0 \arrowvert D \arrowvert \Phi_0 \rangle \nonumber \\
 &=& \langle \Phi_0 \arrowvert e^{T \dag} D_N e^T \arrowvert \Phi_0 \rangle_C + \langle \Phi_0 \arrowvert (-\sum_i e \bm{r_i} + \sum_A Z_A e \bm{r_A}) \arrowvert \Phi_0 \rangle \nonumber \\
 &=& \langle \Phi_0 \arrowvert e^{T \dag} D_N e^T \arrowvert \Phi_0 \rangle_C + \langle \Phi_0 \arrowvert (-\sum_i e \bm{r_i}) \arrowvert \Phi_0 \rangle \nonumber \\
 &+& \sum_A Z_A e\bm{r_A} \langle \Phi_0 \arrowvert \Phi_0 \rangle \nonumber \\
 &=& \langle \Phi_0 \arrowvert e^{T \dag} D_N e^T \arrowvert \Phi_0 \rangle_C + \langle \Phi_0 \arrowvert (-\sum_i e \bm{r_i}) \arrowvert \Phi_0 \rangle \nonumber \\
 &+& \sum_A Z_A e\bm{r_A}
\end{eqnarray}

where $\arrowvert \psi \rangle$ is the electronic wavefunction of the molecule, 
which is expressed as $e^T \arrowvert \Phi_0 \rangle$, in the coupled cluster method. 
$\arrowvert \Phi_0 \rangle$ is the model state, the Dirac-Fock (DF) wavefunction of the ground state of the molecule, which is built from 
single particle four-component spinors. 
T is the cluster operator. In the coupled cluster singles and doubles (CCSD) approximation, which we work with,  
$T = T_1 + T_2$, where $T_1$ and $T_2$ are the single and double excitation operators respectively. 
D is the electric dipole moment operator, e is the charge of the electron, 
summation over the electronic coordinates is indicated by i, and that over the nuclear coordinates by A. $\bf{r_i}$ is the position vector 
from the origin to the coordinate of an electron, and $\bf{r_A}$ is the position vector from the origin to the coordinate of a nucleus. $Z_A$ is 
the atomic number of the $A^{th}$ nucleus. The subscript `C' means that each term in that expression is connected\cite{Kvas, Bishop}, and 
`N' refers to the normal ordered 
form of that operator~\cite{Lindgren}. 
Note that we have invoked the Born-Oppenheimer approximation in the fifth line of the equations given above. 

The important aspects of our relativistic CCSD method are that we use the Dirac-Coulomb 
Hamiltonian, correlation effects have been taken into account to all orders in the residual Coulomb interaction for the one and 
two hole-particle excitations. The coupled cluster method is size extensive, unlike the truncated configuration interaction (CI) method~\cite{Das}.

For all the molecules considered in the present work, the origin is chosen to be the flourine atom, and hence the PDMs can be expressed as  

\begin{eqnarray}
 d &=& \langle \Phi_0 \arrowvert e^{T \dag} D_N e^T \arrowvert \Phi_0 \rangle_C \nonumber \\
 &+& \langle \Phi_0 \arrowvert (-\sum_i e \bm{r_i}) \arrowvert \Phi_0 \rangle +  Z_{A} e r_e  
\end{eqnarray}

where $r_e$ refers to the equilibrium bond length for the molecule AF, with A$=$Be, Mg, Ca, Sr or Ba. 
The first term captures the electron correlation effects, while the second is the electronic contribution from the DF calculations. 
The third gives the nuclear contribution.
The PDM depends on the mixing of orbitals of opposite parity. This is naturally achieved in polar molecules, as their orbitals are hybridized. 

\section{Methodology}
The molecular PDMs in the present work  were calculated by combining the well known UTChem and DIRAC08 codes~\cite{utchem,utchemt,dirac}.  
The DF calculations to generate the orbitals at the Self Consistent Field (SCF) level and the atomic orbital to molecular 
orbital integral transformations ~\cite{abe} 
were carried out using the 
UTChem code. The $C_8$ double group symmetry was used to reduce the computational cost~\cite{c8}. 
The CCSD calculations were carried out in the DIRAC08 code, 
using the one and two electron integrals from UTChem. The electronic part of the PDM 
was calculated by using only the linear terms in the coupled cluster wavefunction, since their contributions are the largest~\cite{Mabe}

\begin{eqnarray}
 \langle\Phi_0\arrowvert(1+T_1+T_2)^\dag(-\sum_i e \bm{r_i})_N(1 + T_1 + T_2)\arrowvert\Phi_0\rangle_C \nonumber \\
 + \langle \Phi_0 \arrowvert (-\sum_i e \bm{r_i}) \arrowvert \Phi_0 \rangle
\end{eqnarray}

In the above expression, the cluster amplitudes are obtained by solving the full CCSD equations containing the linear and the nonlinear terms.
We add the nuclear contribution to the electronic part of the PDM using the experimental value for the bond length, wherever available. 
The values of the bond lengths used for BeF, MgF, CaF, SrF and BaF are 1.361, 1.75, 1.967, 2.075~\cite{Langhoff,Herzberg} and 
2.16~\cite{Mestdagh,Ryz} Angstrom respectively. 

The details of the basis sets used for our computations are given below in Table I. We used uncontracted Gaussian type basis sets (GTOs) 
in all our calculations. We also imposed the kinetic balance~\cite{Dyall} condition for all the basis sets. 
For Sr and Ba, we used the
exponential parameters taken from the four-component
basis sets obtained by Dyall~\cite{Dyallbasis}, and added to it diffuse and polarization functions from the Sapporo-DKH3~\cite{Sap}
basis sets. We used the exponential parameters of
cc-pV (correlation consistent-polarized valence) basis sets
from the EMSL Basis Set Exchange Library~\cite{basissetlibrary,basissetlibraryt} for
Be, Mg, Ca and F. 

\begin{table}[h!] 
 \centering
 \caption{Details of the basis sets used}
 \label{}
 \begin{tabular}{|r|r|}
 \hline
 Atom&Basis \\
 \hline
    Be&cc-pVDZ: 9s, 4p, 1d \\
      &cc-pVTZ: 11s, 5p, 2d, 1f \\
      &cc-pVQZ: 12s, 6p, 3d, 2f, 1g \\
      \hline
    Mg&cc-pVDZ: 12s, 8p, 1d \\
      &cc-pVTZ: 15s, 10p, 2d, 1f \\
      &cc-pVQZ: 16s, 12p, 3d, 2f, 1g \\
      \hline
    Ca&cc-pVDZ: 14s, 11p, 5d \\
      &cc-pVTZ: 20s, 14p, 6d, 1f \\
      &cc-pVQZ: 22s, 16p, 7d, 2f, 1g \\
      \hline
    Sr&Dyall+Sapporo: 20s, 14p, 9d \\
      &Dyall+Sapporo: 28s, 20p, 13d, 2f \\
      &Dyall+Sapporo: 33s, 25p, 15d, 4f, 2g \\
      \hline
    Ba&Dyall+Sapporo: 25s, 19p, 13d \\
      &Dyall+Sapporo: 31s, 25p, 15d, 2f \\
      &Dyall+Sapporo: 36s, 30p, 18d \\
      \hline
     F&cc-pVDZ: 9s, 4p, 1d \\
      &cc-pVTZ: 10s, 5p, 2d, 1f \\
      &cc-pVQZ: 12s, 6p, 3d, 2f, 1g \\
 \hline
 \end{tabular}
\end{table}

\section{Results and Discussions}
Table II gives the results of our calculations of energies of the molecules and their PDMs at DF and CCSD levels. 
The values for the PDMs have been rounded off to the second decimal place. 

\begin{table}[h] 
 \centering
 \caption{Summary of the calculated results of the present work}
 \label{}
 \begin{tabular}{|r|r|r|r|r|}
 \hline
 $Molecule$ &$Method$ &$Basis$ &$E (au)$&$PDM (D)$  \\
 \hline
  BeF&DF&cc-pVDZ &-114.07 &1.32   \\
     &DF&cc-pVTZ &-114.23 &1.31  \\
     &DF&cc-pVQZ &-114.26 &1.30  \\
     &CCSD&cc-pVDZ &-114.38 &0.93  \\
     &CCSD&cc-pVTZ &-114.59 &1.06  \\
     &CCSD&cc-pVQZ &-114.67 &1.10  \\
     &Expt&- &- &-  \\
 \hline
  MgF&DF&cc-pVDZ &-299.51 &3.21   \\
     &DF&cc-pVTZ &-299.52 &3.21  \\
     &DF&cc-pVQZ &-299.57 &3.16  \\
     &CCSD&cc-pVDZ &-299.96 &2.84  \\
     &CCSD&cc-pVTZ &-300.02 &3.02  \\
     &CCSD&cc-pVQZ &-300.11 &3.07  \\
     &Expt&- &- &-  \\
 \hline
  CaF&DF&cc-pVDZ &-779.31 &2.89   \\
     &DF&cc-pVTZ &-779.33 &2.82  \\
     &DF&cc-pVQZ &-779.37 &2.77  \\
     &CCSD&cc-pVDZ &-780.09 &3.01  \\
     &CCSD&cc-pVTZ &-780.21 &3.13  \\
     &CCSD&cc-pVQZ &-780.31 &3.16  \\
     &Expt&- &- &$3.07(7)$  \\
 \hline
  SrF&DF&cc-pVDZ &-3277.67 &2.83   \\
     &DF&cc-pVTZ &-3277.70 &2.95  \\
     &DF&cc-pVQZ &-3277.74 &3.01  \\
     &CCSD&cc-pVDZ &-3278.85 &2.95  \\
     &CCSD&cc-pVTZ &-3279.01 &3.42  \\
     &CCSD&cc-pVQZ &-3279.13 &3.60  \\
     &Expt&- &- &$3.4676(1)$  \\
 \hline
  BaF&DF&cc-pVDZ &-8235.25 &2.42   \\
     &DF&cc-pVTZ &-8235.27 &2.28  \\
     &DF&cc-pVQZ &-8235.31 & 2.65 \\
     &CCSD&cc-pVDZ &-8236.55 &2.69  \\
     &CCSD&cc-pVTZ &-8236.71 &3  \\
     &CCSD&cc-pVQZ &-8236.82 & 3.40 \\
     &Expt&- &- &$3.170(3)$  \\
 \hline
 \end{tabular}
\end{table}

We observe that the absolute value of the correlation energy increases as the molecules get heavier, that is, 
as the relativistic effects get more pronounced. 
In each of the molecules BeF, MgF, and CaF, the PDM decreases at the 
DF level as we move from DZ through the QZ basis sets, while for BaF,  
the PDM oscillates. 
For a given molecule, the CCSD values of this quantity increase progressively as the size of the basis set is enlarged.
Also, the CCSD results for the PDM converge, that is, 
the difference between the QZ and the TZ basis set PDMs are less than that between TZ and DZ, except for BaF. 
This may either be due to the correlation being inadequate or 
due to insufficient optimization of the basis sets. 
The PDM increases from BeF to SrF, that is, the PDM increases as a system becomes more relativistic, but drops slightly for BaF.  
We also see that the absolute value of the correlation effects monotonically increase as we go to more relativistic systems, with MgF 
being an exception. 
The effect of correlation is 
0.2 D for BeF, whereas it is 0.75 for BaF. 
The $T_1$ diagonostics, for all the CCSD calculations and for all 
the basis sets used, was about 0.02 
for all the molecules, except BaF, for which it was 0.01.  
This is an indicator of the stability of our single reference calculations. 
Our calculated values of the PDM at the QZ level differ from experiment by about 3, 4 and 7 percent  for CaF, SrF and BaF respectively. 

We tabulate the electronic (at the CCSD level, with QZ basis) and nuclear contribution to the PDM for the 5 molecules in Table III. 

\begin{table}[h] 
 \centering
 \caption{Electronic (at the CCSD level, with QZ basis sets) and nuclear contributions to the PDM for all the alkaline earth monofluorides}
 \label{}
 \begin{tabular}{|r|r|r|}
 \hline
  Molecule &Electronic term&Nuclear term\\
 \hline
  BeF&-25.05&26.15  \\
  MgF&-97.80&100.87  \\
  CaF&-185.81&188.97  \\
  SrF&-375.16&378.76  \\
  BaF&-577.64&581.04  \\
 \hline
 \end{tabular}
\end{table}

We see from Table III that the absolute values of the electronic and the nuclear terms increase with the size of the molecules. 
The nuclear contribution, which is positive, `overtakes' the electronic 
contribution, which is negative, as the  
molecules get heavier, up to SrF. 

We now consider the ratio of the electronic terms to the nuclear terms. Their absolute values from
BeF to BaF, 0.96, 0.97, 0.98, 0.99 and 0.99 respectively. The ratio for BaF is higher than that for SrF, with the difference occurring in 
the third decimal place. 

To understand correlation trends, we consider the contributions from each of the 9 terms in Eq. 3. 

\begin{table}[h] 
 \centering
 \caption{Contributions from the individual terms to the PDM for all the alkaline earth monofluorides}
 \label{}
 \begin{tabular}{|r|r|r|r|r|r|}
 \hline
  Term &BeF&MgF&CaF&SrF&BaF\\
 \hline
  DF&-24.85&-97.72&-186.20&-375.75&-578.39  \\
  D$T_1$&-0.08&-0.02&0.21&0.31&0.4  \\
  D$T_2$ &0&0&0&0&0 \\
  $T_1^\dag D$&-0.08&-0.02&0.21&0.31&0.4  \\
  $T_1^{\dag} D T_1$ &-0.02&-0.02&-0.02&-0.03&-0.05 \\
  $T_1^{\dag} D T_2$ &0.02&0.01&0.02&0.02&0.02 \\
  $T_2^{\dag} D$ &0&0&0&0&0 \\
  $T_2^{\dag} D T_1$ &0.02&0.01&0.02&0.02&0.02 \\
  $T_2^{\dag} D T_2$ &-0.06&-0.04&-0.05&-0.04&-0.04 \\ 
 \hline
 \end{tabular}
\end{table}

We observe that the largest contribution to the PDM comes from the DF term. The D$T_1$ term, which embodies important pair correlation effects~\cite{Eliav} 
contributes the most, among the correlation terms, in the 
heavier molecules, starting from CaF, and increases from CaF to BaF. It is, however, not
significant for BeF and MgF.  The D$T_1$ term steadily increases from BeF to BaF. Also, the term changes sign as it increases with the size of the molecule. 
Furthermore there are cancellations between the various correlation effects. 
The fractional contributions of the correlation effects increase from lighter to heavier elements, except  in the case of BeF. It is largest in the case of BaF for which
the value is 0.22.

We have given below the comparison 
of our results with previous calculations and experiment. 

\begin{table}[h] 
 \centering
 \caption{Comparison of present work with previous calculations and experiment}
 \label{}
 \begin{tabular}{|r|r|r|r|}
 \hline
 $Molecule$&$Work$&$Method$&$PDM (D)$  \\
 \hline
  BeF&         Langhoff et al~\cite{Langhoff}&CPF&1.086   \\
         &        Buckingham et al~\cite{Buckingham}&MP2&1.197               \\
      &         Kobus et al~\cite{Kobus}&FD-HF&-1.2727               \\
       &This work (QZ)&&1.10                \\
       &     Expt&-&-  \\
 \hline
  MgF&         Torring et al~\cite{Torring}&Ionic model&3.64   \\
        &       Langhoff el al~\cite{Langhoff}&CPF&3.077             \\
        &       Mestdagh el al~\cite{Mestdagh}&EPM&3.5\\
        &       Buckingham el al~\cite{Buckingham}&MP2&3.186\\
        &       Kobus el al~\cite{Kobus}&FD-HF& -3.1005            \\        
       &This work&   & 3.06         \\
        &    Expt&-&-  \\
 \hline
 
       CaF  &      Torring et al~\cite{Torring}&Ionic model&3.34           \\
       &      Rice et al~\cite{Rice}&LFA&3.01\\
         &      Langhoff et al~\cite{Langhoff}&CPF&3.06\\
         &      Mestdagh et al~\cite{Mestdagh}&EPM&3.2\\
          &      Bundgen et al~\cite{Bundgen}&MRCI&3.01\\
           &         Allouche et al~\cite{Allouche}&LFA&3.55   \\
            &      Buckingham et al~\cite{Buckingham}&MP2&3.19\\
         &      Kobus et al~\cite{Kobus}&FD-HF&-2.6450\\
       &This work&&   3.16         \\ 
        &    Expt~\cite{Childs}&&$3.07(7)$  \\
 \hline
       SrF  &      Torring et al~\cite{Torring}&Ionic model&3.67         \\
         &      Langhoff et al~\cite{Langhoff}&CPF&3.199\\
         &      Mestdagh et al~\cite{Mestdagh}&EPM&3.59\\
         &         Allouche et al~\cite{Allouche}&LFA&3.79   \\
         &      Kobus et al~\cite{Kobus}&FD-HF&-2.5759\\
         &Prasannaa et al~\cite{VSP}&CCSD&3.41 \\
         &Sasmal et al~\cite{Sudip}&Z-vector&3.4504 \\
       &This work&  & 3.60     \\ 
        &    Expt~\cite{Kandler}&&$3.4676(1)$  \\
 \hline
        BaF &      Torring et al~\cite{Torring}&Ionic model&3.44        \\
          &     Mestdagh et al~\cite{Mestdagh}&EPM&3.4\\
           &         Allouche et al~\cite{Allouche}&LFA&3.91   \\
       &This work&   &3.40    \\
        &    Expt~\cite{Ernst}&&$3.170(3)$  \\
 \hline
 \end{tabular}
\end{table}

The first calculations on the PDM of some of the alkaline earth monofluorides were carried out by Torring et al~\cite{Torring}. They used an ionic model to calculate 
the PDMs of MgF, CaF, SrF and BaF. 
They compare their results with the Rittner model~\cite{Rittner} and experiment (wherever available) in their work. 
Childs et al experimentally determined the PDM of CaF to be 3.07(7) D in their work~\cite{Childs} in the same year. 
In 1985, the PDM of SrF was measured, using a molecular beam microwave double resonance method~\cite{Kandler}. 
Rice et al~\cite{Rice} used the Ligand field approach (LFA) to obtain a value of 3.01 D for CaF, later that year. 
The PDM of BaF was measured to be 3.170 (3) by Ernst et al in the subsequent year~\cite{Ernst}. 
The first ab initio calculations on select alkaline earth monofluorides were performed by Langhoff et al~\cite{Langhoff}. 
They employed extended Slater basis sets augmented with diffuse and polarization functions. They calculated the 
PDMs using 
the single reference configuration interaction singles and doubles (CISD) and the coupled pair functionals (CPF) methods. 
The CISD method does not take into account certain higher order correlation effects that are present in the CCSD method.
The CPF approach is a size consistent version of CISD, but the treatment 
is non-relativistic. Also, not all the electrons are correlated in these calculations. 
Mestdagh et al~\cite{Mestdagh} used an electrostatic polarization model (EPM) for calculating the PDMs of MgF, CaF, SrF and BaF, among other molecules. 
Bundgen et al~\cite{Bundgen} employed the multireference-configuration interaction (MRCI) approach, with the singles and doubles excitations taken into account, 
to compute the PDM of CaF. 
They used Gaussian basis sets, and a total of 17 electrons were correlated in 
their calculations. 
Allouche et al~\cite{Allouche} used the LFA, in 1993, to calculate the PDMs of Ca, Sr and Ba monohalides. 
Buckingham et al~\cite{Buckingham} 
used Second order Moller-Plesset perturbation theory (MP2) to calculate PDMs of BeF, MgF and CaF. In CCSD, the single and double
excitations to all orders of perturbation are taken care of. So, we believe that CCSD can capture correlations better at a finer level. 
Also, their calculations are non-relativistic. 
The work by Kobus et al~\cite{Kobus} was to compare the electric moments obtained from finite basis set with finite difference 
Hartree-Fock (FD-HF) calculations. 
In an earlier work~\cite{VSP}, we had tested the accuracy of our RCCM by computing the PDM of the SrF molecule. 
Sasmal et al~\cite{Sudip} improved upon our result of SrF in their work, which involved implementing the Z vector approach in a RCCM. 

The possible sources of errors in our calculations stem from the higher order terms neglected, and the choice of basis sets. 
If we add up the individual terms that make the PDM (other than the DF, D$T_1$ and its conjugate), we see that the highest sum is 
from BaF, which is -0.05 D. It is reasonable to assume that the neglected higher order terms do not exceed this value, and we set a conservative 
estimate of $\pm$0.1 D. 

The error from the choice of basis sets can be estimated by taking the difference between the QZ and the TZ PDMs. They are 
0.04, 0.05, 0.03 and 0.18 for the first four alkaline earth monofluorides. We set a conservative error of $\pm$0.1 D for BeF, 
MgF and CaF, and $\pm$0.2 D for SrF. 
The PDM doesn't converge at the QZ level for BaF, so 
we cannot determine the error due to incompleteness of the basis here. 
However, we can roughly estimate the error to be about 0.2 D, based on comparison with a preliminary investigation that has been carried out using a larger basis based on the same method that we have used in our calculations ~\cite{Sunaga}.

For SrF and BaF, we have used a combination of 
Dyall's and Sapporo's basis for Sr and Ba. These contain in them diffuse and polarization functions, just as Be, Mg and Ca had 
these functions via cc-pV. In order to see the trends across the alkaline earth monofluorides, we have to assume that 
the results do not change significantly, whether we use cc-pV basis or a combination of Dyall and Sapporo basis. The error from this 
assumption is tested using CaF at the QZ level, since Ca is the only atom in the candidate molecules for which both Dyall and cc-pV basis sets were available. 
We obtain a PDM of 2.84 and 3.23 D at the DF and the CCSD levels, respectively, when we use Dyall plus Sapporo basis for CaF. Earlier, we obtained, 
using cc-pVQZ for Ca, 2.77 and 3.16 D at the DF and CCSD levels, respectively. The difference is about 2.5 percent at DF  and 2.2 at CCSD. 
Since we can't tell how this may vary for the heavier molecules, we can conservatively set an error percentage of about 5, due to the change in 
the choice of basis sets for the molecules. 

\section{Conclusions}
We calculated the PDMs of the alkaline earth monofluorides, up to BaF, using a RCCM, with no core orbitals frozen in our calculations. 
We used uncontracted cc-pV and Dyall+Saporro basis sets for our calculations. We reported the DF and CCSD energies and the PDMs. 
Our results, using QZ basis sets, are in good agreement 
with the experimental results, wherever available. We have examined  the electronic and nuclear contributions to the PDMs for all the molecules considered, 
as well as the importance of the individual correlation terms. 
The results we obtained suggest that as the molecules get heavier, that is, as the relativistic effects increase, the correlation effects get larger in size. 
We also provided a rough estimate of the errors in our calculations, caused by ignoring higher order terms and from the basis sets. We also give a rough 
estimate of the error in comparing the PDMs as we move from BeF through BaF, due to the choice of basis sets being different for the two heaviest 
monofluorides. 
 
\section{Acknowledgment}

The computational results reported in this work 
were performed on the high performance computing facilities of IIA, Bangalore, on the hydra and kaspar clusters. We would like to acknowledge 
the help of Anish Parwage, Engineer C, Indian Institute of Astrophysics, Bangalore, for his help in the installation of the codes on the clusters. 
The DiRef database was extremely useful in 
searching for literature~\cite{diref}.

\end{document}